# Direct observation of the crossed interhemispheric transfer of the left-right mirror-images in human vision

Albert Le Floch [a,b] , Guy Ropars [a,c*]

[a] *Laser Physics Laboratory, University of Rennes, 35042 Rennes Cedex, France*

[b] *Quantum Electronics and Chiralities Laboratory, 20 Square Marcel Bouget, 35700 Rennes, France*

[c] *UFR SPM, University of Rennes, 35042 Rennes Cedex, France*

*\* Corresponding author at: Laser Physics Laboratory, University of Rennes, 35042 Rennes Cedex, France. e-mail: guy.ropars@univ-rennes.fr ; albert.lefloch@laposte.net*

Symmetry breaking is common in animal and human brains where the lack of asymmetry often perturbs behavioral and cognitive functions. In particular, the ubiquity of mirror-image confusion in young children, which often persists in dyslexia, is established. However, the very existence of these symmetric mirror-images and their perceptual or memory nature remain controversial. Here, using the noise-activated afterimage method, we demonstrate that a dyslexic with mirror-images is an ideal candidate for solving the mystery. Indeed, after a monocular fixation, the primary afterimages are perceived alone through this eye, while the mirror-images are also perceived alone, but exclusively through the other eye which has remained closed. We deduce that the callosal interhemispheric connections are necessarily projected on the dominance columns of layer 4 of the primary cortex, the only layer where segregation is strict, and are furthermore crossed. Our results show that perceived primary and mirror images are spatially and temporally resolved.

# I- Introduction

Since Ernst Mach noted at the end of the 19th century[1] that young children constantly confound symmetric letters like b and d, or p and q, many studies have been published since the 20th century on mirror-reversal errors of letters in reading, copying and writing. The large body of literature on the problem has been reviewed in refs[2–4] examining the different brain theories which have been proposed, and synthetizing the recent works and ideas.

Early experiments with animals like pigeons[5] and monkeys[6–8] have also shown their great difficulty in learning to discriminate pairs of mirror-images, confirming the ubiquity of mirror confusion. These studies suggest that this left-right problem is related to the structural symmetry of the nervous systems. It has recently been shown[9] that the left-right asymmetry of the visual system in drosophila controls its orientation toward a visual object. In addition, greater asymmetry corresponds to better orientation. Furthermore, the experiments in species with corpus callosum[10], and studies of split-brain patients[11] have demonstrated the crucial role played by the corpus callosum in interconnecting the two brain hemispheres and cementing together the two halves of the visual field[12]. In animals, mirror transfer is indeed lost when the interhemispheric bundles of fibres and specifically the corpus callosum are cut[6,13,14]. This work suggests that in humans, some asymmetry is required in specific tasks like reading and writing. Mach himself[1] suggested that although a symmetry underlies mirror-image confusion in humans, a slight asymmetry somewhere in the brain is required to distinguish between primary and secondary mirror images such as b and d. As noted by many authors [2,4,15,16], these mirror-image confusions are also often reported in persons with dyslexia. In contrast to typical observers, in such observers a lack of asymmetry between the two retinal Maxwell centroids, i.e. the photoreceptor topographies, has been shown to perturb the interhemispheric connections[17]. Indeed, while children without dyslexia are able to eliminate the mirror-image of letters at 5-to-6 years old[18], the confusion persists for some children with dyslexia. In a cohort of 160 children and teenagers with dyslexia, the mirror-images are perceived by 60% of them[19]. When b is used as a stimulus in binocular fixation for instance, they see the two superposed noise-activated afterimages b and d.

The debate attempting to explain the letter reversals and confusions in humans was initiated by Orton[15,16]. In this model, due to symmetry, the real image is recorded by one hemisphere, while a reversed memory mirror-image is directly laid down in the other hemisphere. Mirror-image confusion would result from failure to develop cerebral dominance, so as to suppress the reverse memory. However, the model is too simplistic, in particular as a real image on the retina itself is split between the two hemispheres through the optical chiasm for each eye. Nevertheless, the model introduces the concept of mirror-image linked to symmetry which leads to mirror errors in reading and writing for many children. A different model for the interhemispheric mirror-image reversal was proposed after experiments with monkeys[6,7] requiring a transfer of information essentially through the corpus callosum linking the two hemispheres. In this model, both hemispheres are supposed to perceive the primary as well as the mirror images. The necessary integration of the two half-fields for the primary image seems to conflict with the integration of the two perceived half-fields of the mirror-image via homotopic interhemispheric connections[4]. To try to avoid this difficulty, a process of homotopic mapping between the hemispheres of the brain was suggested[14] by introducing only a memory reversal in the higher levels of the brain, instead of the perceived reversal on the primary cortex. The interhemispheric transfer of the primary image is perceptual but the transfer of the mirror-image requires a memory registration. However, the nature and the loci of the projections are not identified in any model, and moreover, in the last model, the brain is supposed to be perfectly symmetric at the memory level.

We may wonder if the superposed perception of the primary and secondary mirror-image by an observer with dyslexia seen after a binocular fixation[17] using the noise-activated afterimage method that we extend here, can bring new insights into the basic mechanisms governing interhemispheric connections, in their loci and their role in the human brain. In contrast to different other species[20], in humans and in some primates such as macaques[21] the corpus callosum plays a specific role[11,22–24] in the architecture of laminar visual connections in primary cortex V1, notably in layer 4 [21]. While the non-symmetric stitching connections establish the continuity of the visual field, the symmetric connections link points of the visual field following the principle of symmetry[4,20,25].

It is the aim of our study to investigate the nature and the exact loci of the symmetric interhemispheric mirror image projections, with the help of a young adult CT with dyslexia who sees mirror-images, by using the noise-activated afterimage method extended here to monocular fixations. We use the same method to explore the brain basis for the transfer of letter, bigram, word and non-word reversals which can all induce an internal visual crowding during the learning to read process. Using monocular fixation, we can identify the crossed nature of these symmetric interhemispheric projections occurring alone between the respective dominance columns[26,27] of the two eyes. As the primary and mirror-image are then observed separately by admitting the noise alternatively in only one eye, we have to define the loci of the projections in the layer 4 of the primary cortex V1, the only layer where the strict segregation of the dominance columns of each eye is preserved[12,28]. Moreover, as the transfer through the corpus callosum implies an unavoidable small delay[29,30], we suggest that the primary and mirror-images then appear both spatially and temporally resolved. This specific connectivity of the callosal projections will give us the possibility of weakening the disturbing extra mirror-images for readers with dyslexia who lack asymmetry, using Hebbian mechanisms[31]. Moreover, this crossed architecture can also be an interesting point of investigation for the transfer of the retinal Maxwell centroids asymmetry to the primary cortex. This could explain the usual weakening during the development of the mirror-images in children with a normal reading status at the end of the critical period for their visual system, i.e. at about 5 to 6 years old during which the neuronal receptive fields have been shown to be sculpted[32]. Finally, we will discuss the crucial role of the presence or absence of asymmetry in the development of different visual pathways.

## II- Results

*1-The lack of asymmetry between the Maxwell's centroids*

The photoreceptors in human retinas are not regularly arranged on straight lines like the red, green and blue pixels on a computer screen. In contrast, around the centre of the fovea, the cone mosaic shows a rather concentric distribution (Fig. 1a). In particular, the blue cone migration occurring as soon as 20 week gestation, before the fovea begins to form[33], has evolved towards a specific adult-like blue cone-free area of about 100 μm in diameter at the centre of the fovea. The existence of this blue cone-free area has first been inferred indirectly from the physiology of vision[34], then directly observed on retina post-mortem after staining the blue cones with a specific antibody[35]. This area plays a crucial role in human vision acuity, the foveal pit being devoid of all layers of the retina and the chromatic dispersion being canceled for the blue part of the spectrum[36]. Moreover, it corresponds to the Maxwell centroid of the so-called Maxwell spot[37] whose outlines can be recorded using a foveascope (see Methods). In general, the ellipticity of the two outlines are different for a typical observer (see the Supplementary Fig. S1). The more circular outline defines the dominant eye[17]. Here the Maxwell centroid outlines for CT are shown in Fig. 1b. The ellipticities of both eyes are similar and equal to about 1, as the centroids are quasi-circular. The diameter of both centroids is $29' \pm 3'$ (i.e. $145 \pm 15\ \mu m$).

Moreover, the lack of asymmetry is accompanied by an absence of ocular dominancy for CT, as often observed in dyslexia[38,39].

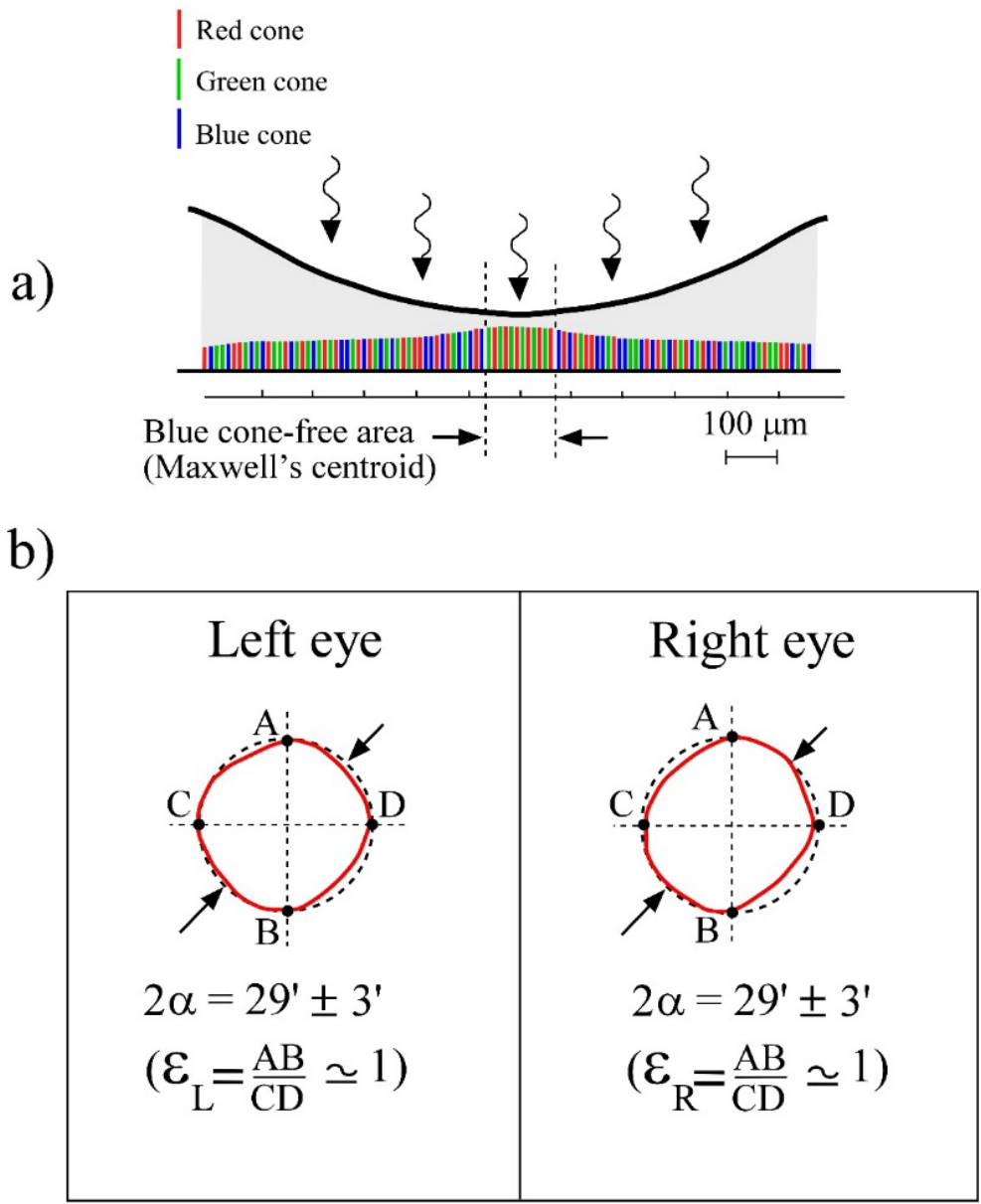

*Figure 1. Maxwell centroids in the foveas for the dyslexic observer.*

a) *Schematic representation of the human fovea with the blue cone-free area at the centre of the cone topography (adapted from Poliak[40])*
b) *The Maxwell's centroids for the CT's two foveas. The two quasi-circular outlines show no noticeable asymmetry ($\Delta\varepsilon = \varepsilon_R - \varepsilon_L \simeq 0$), where $\varepsilon_R$ and $\varepsilon_L$ are the ellipticities of the right and left eye respectively.*

## 2-Noise-activation of an asymmetric double–well quartic potential in a neuron

The behaviour of many bistable nonlinear systems in electronics and sensory biology may be illustrated by the motion of a particle trapped in a symmetric quartic two-well potential of the form[41]:

$$U(x) = -\alpha x^2 + \beta x^4 . \tag{1}$$

The corresponding symmetric scheme, with the two stable states, is represented by the dotted line in Fig. 2a. Neurons involved in the visual pathways in the retina are also bistable and

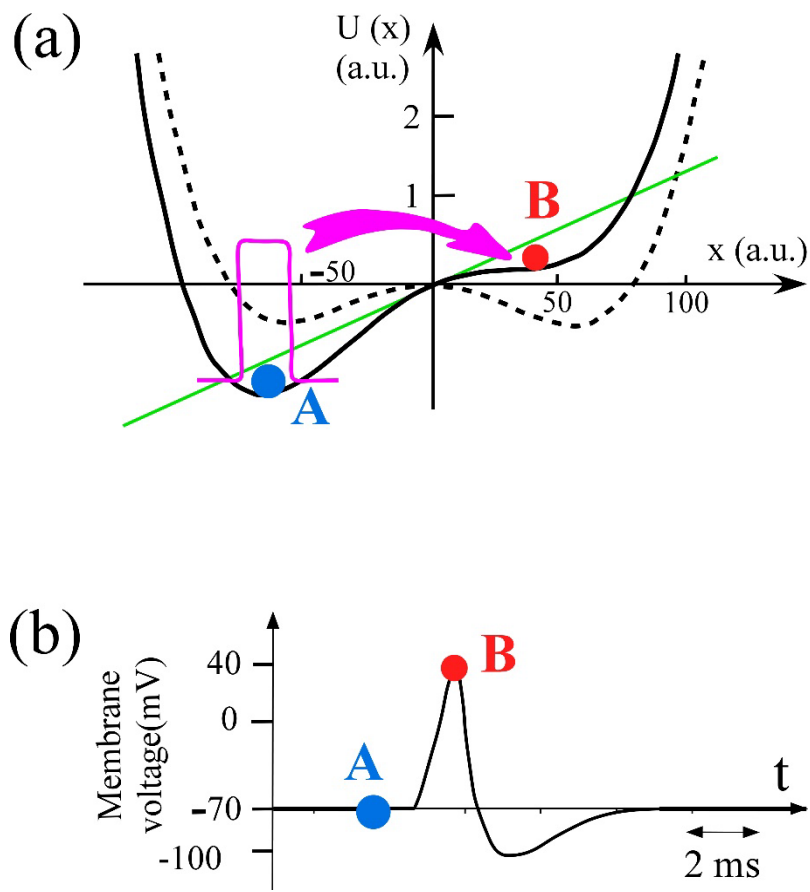

**Figure 2. The asymmetric potential of a neuron membrane.**

a) Scheme of the asymmetrical potential representing the electrochemical potential U(x) of the neural membrane (solid line). The straight green line represents the ionic driving force. The activation signal (in purple) induces the action potential.
b) Typical membrane voltage with its two stable and unstable states during an action potential (A and B respectively).

nonlinear. However, for a neuron, while the resting state is stable, the excited state is unstable (Fig. 2b). To describe the electrochemical potential U(x) of the ganglion cells which generates the action potentials in the retinas, we have to introduce the driving force contributions of the ions through the ionic channels[42,43]. Thus the corresponding asymmetric double-well electrochemical potential (the solid line in Fig. 2a) of a neuron can then be written:

$$U(x) = -\alpha x^2 + \beta x^4 + \gamma x,  \qquad (2)$$

where the order parameter x corresponds to the electrical membrane voltage and $\gamma$ represents the ionic driving force. When the driving force contribution is included in the electrochemical potential U(x) of the neuron membrane, the system can be represented by the asymmetrical potential of Fig. 2a, where the linear term breaks the potential symmetry, as in lasers in the presence of a lever[44]. Hence, we are left with a stable state A corresponding to the resting state of the neuron and an unstable state B corresponding to the peak of the action potential (Fig. 2b). For a neuron at rest, the interior of the cell has a negative voltage of about -70 mV. Usually, when an eye is open, an analog signal (purple line in Fig. 2a) induced by a stimulus in the photoreceptors generates action potentials in a ganglion cell shown in Fig. 2b, following the simplest functional pathway to the brain through the retina which includes photoreceptor cells, bipolar cells and a single ganglion cell[45].

The dynamics of such a visual neural membrane in the presence of an external noise, can then be derived from eq. 2, in the damping regime[46]:

$$\frac{dx}{dt} = -\frac{dU(x)}{dx} = 2\alpha x - 4\beta x^3 - \gamma + \xi(t), \qquad (3)$$

where $\xi(t)$ denotes an external additional zero-mean Gaussian noise impinging on the visual neuroreceptors. In vision this noise can be provided by the diffuse light passing through the eyelids of a closed eye[47] (see below). Noise is usually considered as a nuisance, but in nonlinear systems, noise can enhance the detection of small signals[41]. Moreover, noise can here discriminate between dark and bright parts of a contrasted stimulus as shown in Fig. 3. Indeed, after a fixation, the dark part of the stimulus will correspond to the unbleached photoreceptors that are highly sensitive to the noise (Fig. 3a), while the bright part of the stimulus will correspond to the bleached photoreceptors which appear quasi-insensitive to the same noise falling on the closed eye leading to a dark background (Fig. 3b). Consequently, after a fixation inducing different bleached levels which persist for a few minutes[48], by chopping the noise falling on closed eyes, we can predict that an observer will perceive

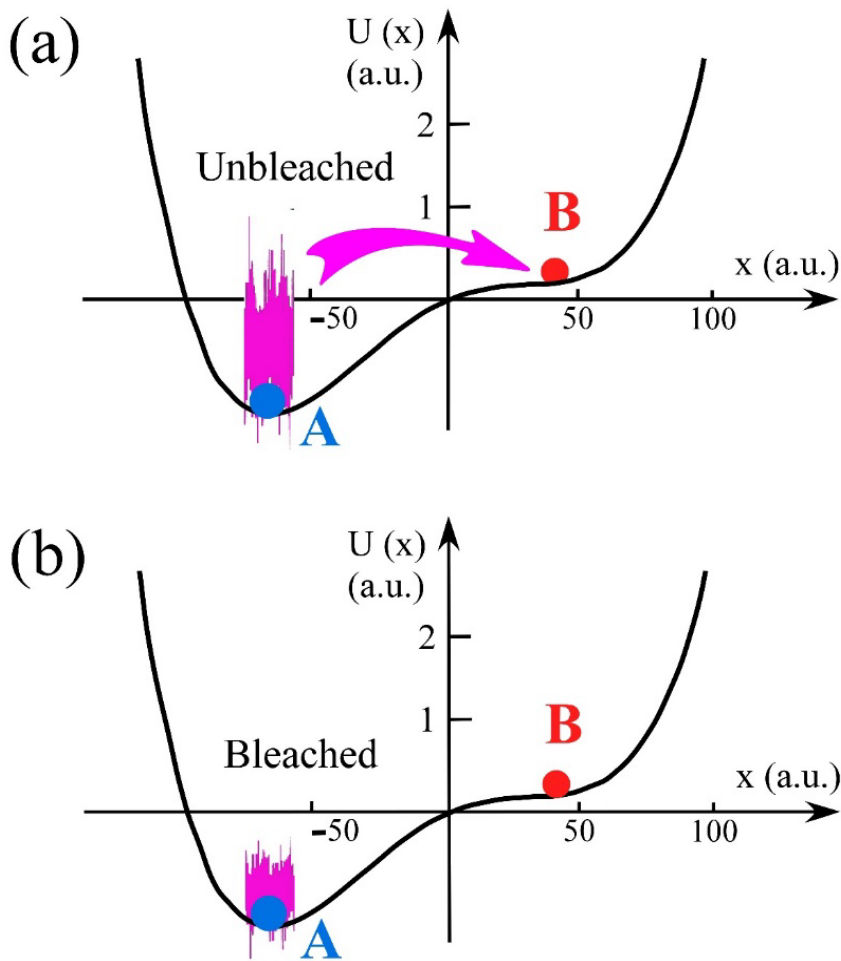

*Figure 3. Principle of the noise activation of afterimages after a differential bleaching.*

a) *Activation of an action potential with noise falling on an unbleached photoreceptor.*
b) *Non activation of action potential with the same noise but falling on a bleached photoreceptor.*

a negative afterimage of a contrasted stimulus (see Methods), the unbleached parts of the stimulus appearing bright while the bleached parts will appear dark.

*3-Noise-activated afterimages of letters after a binocular fixation*

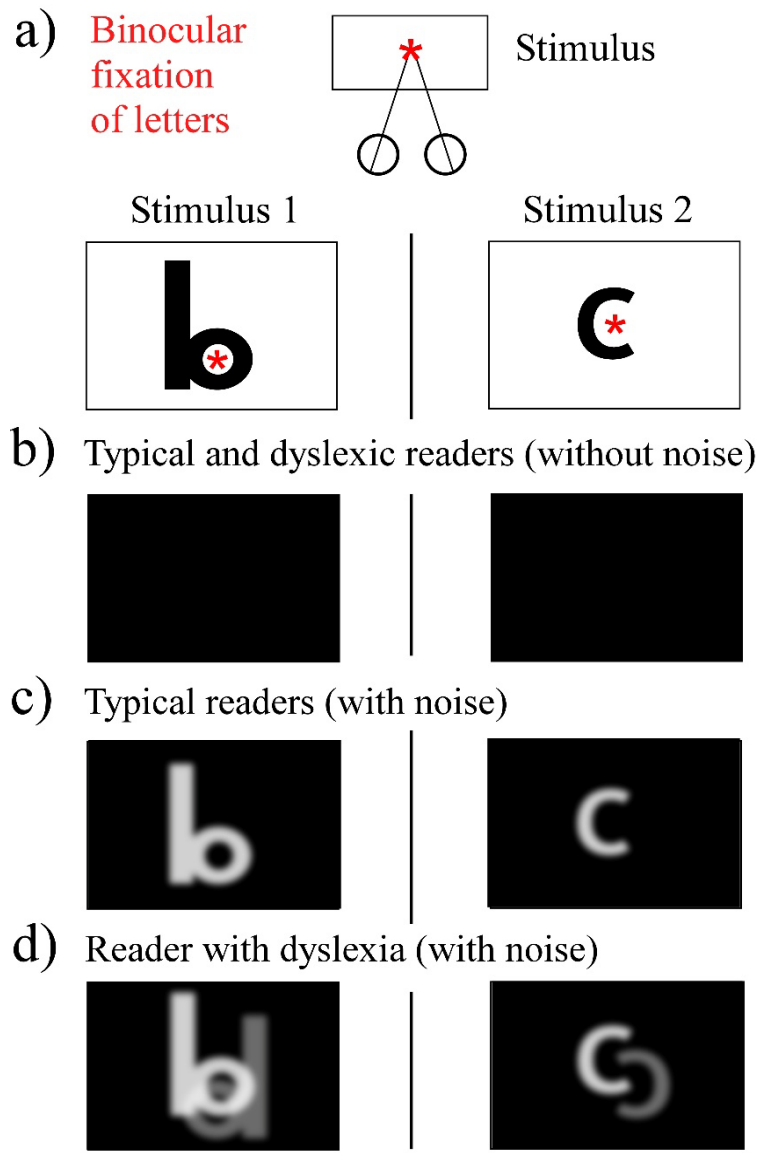

**Figure 4. Binocular fixation of letters.**

a) The two stimuli b and c
b) No afterimage for typical or dyslexic readers in absence of noise.
c) Noise activated afterimages observed by the typical observers.
d) Noise activated afterimages seen by CT. Each letter is accompanied by its mirror image.

---

Using high contrast patterns as stimuli including mirror symmetrical letters and non-mirror letters (Fig. 4a), after a few seconds of binocular fixation, by modulating the noise falling on the eyelids, no afterimage in absence of noise both for typical and for dyslexic readers is first observed (Fig. 4b). In presence of noise, typical readers perceive precise negative primary afterimages for any letters, which are reconstructed in figure 4c. In contrast, CT perceives both the primary and the mirror-images (Fig. 4d). The two images appear superposed with the mirror-image slightly weaker and shifted here toward the right and below the primary image. For all letters, an internal visual crowding effect occurs for CT and moreover for mirror symmetrical letters like b and d, direct confusion is possible.

*4-Noise-activated afterimages of letters and words after a monocular fixation*

When we use a monocular fixation through the left eye (Fig. 5a), the results are surprising. First, for CT, the primary image b is perceived alone due to the noise passing only through the left eyelid, and second the mirror-image d is also perceived alone but when noise passes only through the eyelid of the right eye which has remained closed. As layer 4 received most of the afferents from the lateral geniculate nucleus and is the only layer with strict segregation between the two eyes[12,22], we deduce that the symmetric interhemispheric callosal projections take place between the dominance columns of the two eyes. In Fig. 5d we verify that the same architecture governs the symmetrical projections from the right eye. The connections arriving from each eye on layer 4 thus lead to symmetric interhemispheric projections which are crossed between the dominance columns of the two eyes as schematized in Fig. 6. The superposed primary and mirror images perceived in binocular observation appear now spatially and temporally resolved as the transfer through the callosum is estimated to be about 10 ms [30]. Hence, the crossed internal visual crowding due to the mirror-image is furthermore delayed.

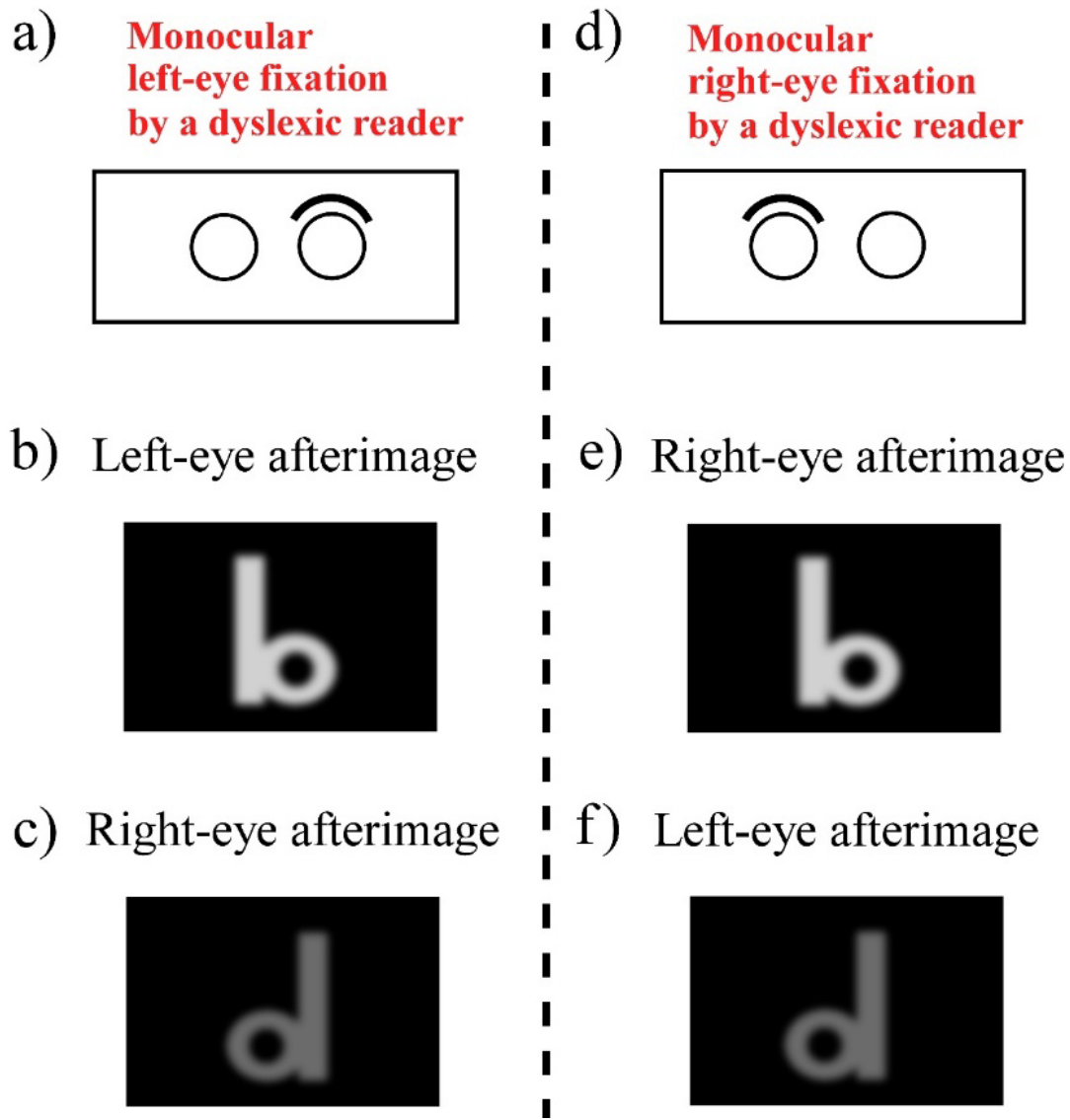

*Figure 5. Monocular fixation of letters by CT.*

a) *Left eye fixation. After this fixation, the two eyelids are closed.*
b) *Noise-activated afterimage in the left eye.*
c) *Noise-activated mirror afterimage in the right eye which has remained closed during the fixation. For each eye, the afterimages in b) and c) are alone since the symmetric interhemispheric projection is crossed between the segregated dominance columns of the two eyes.*

d-e-f) *Similar results after a right eye fixation.*

---

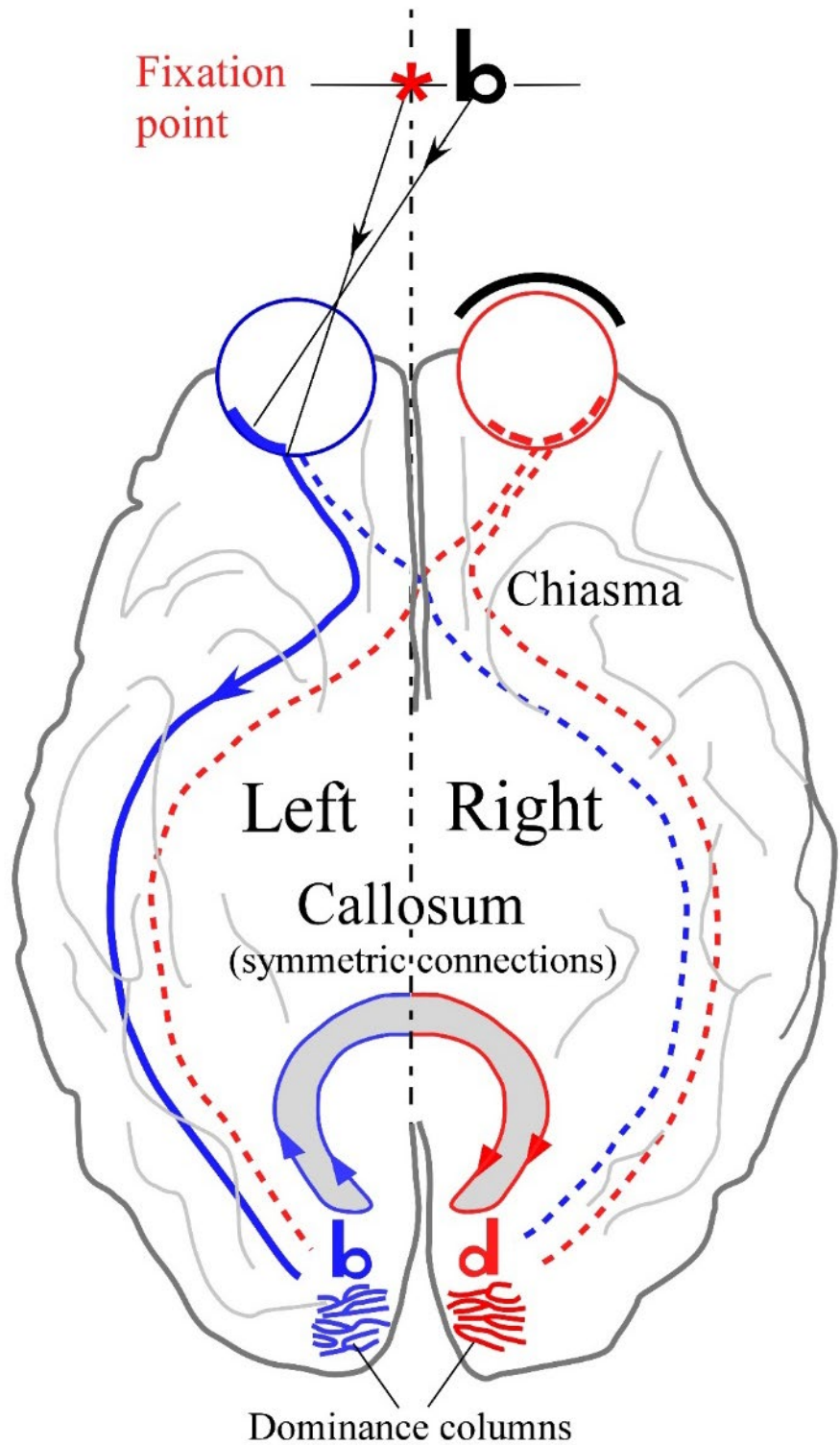

*Figure 6. **Scheme of the neural pathways after a monocular fixation.*** *The primary image of the stimulus b in the right-half field is projected on the segregated dominance columns of layer 4 of the left eye. The crossed symmetric mirror image is projected on the corresponding dominance columns of the right eye.*

Moreover, the reversals are not restricted to individual letters. When we consider a bigram as the stimulus in Fig. 7a, a binocular fixation produces a superposition of primary and mirror images (Fig. 7b). Each monocular fixation (Fig. 7c, d) separates primary and mirror images like for letters, but furthermore with a permutation of the letters of the bigram. For a

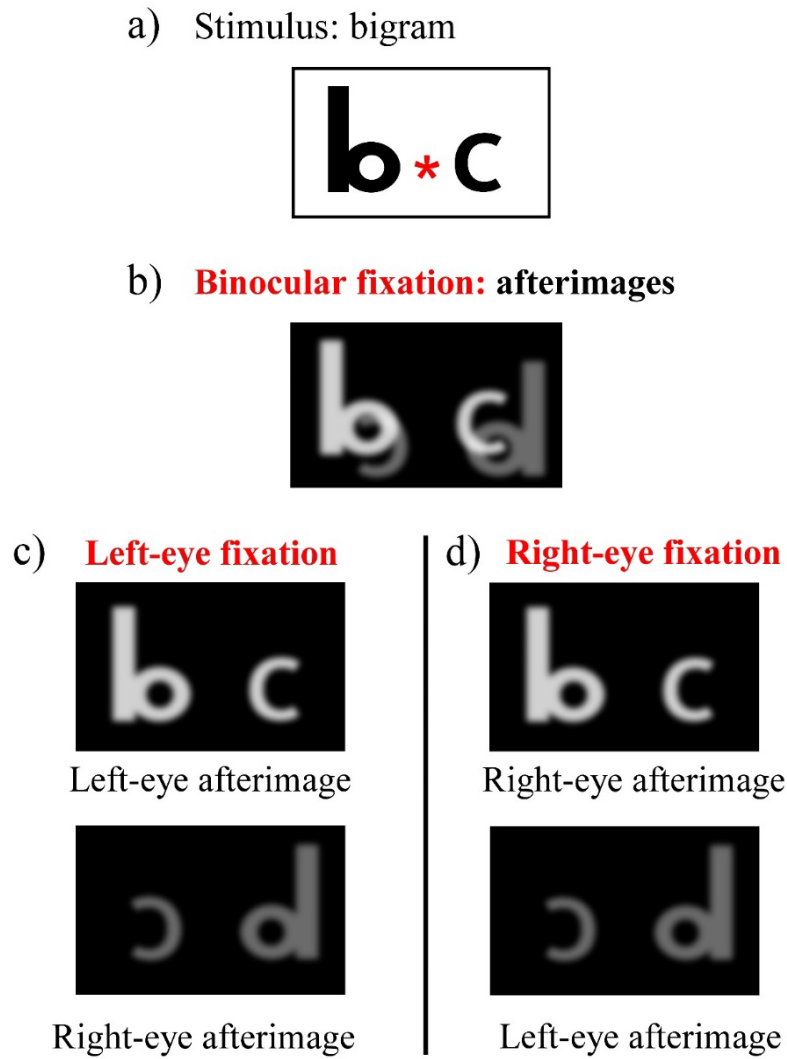

*Figure 7. Bigram noise-activated afterimages observed by CT.*
  a) *The stimulus with the fixation point (red star).*
  b) *Observed noise-activated afterimages after a binocular fixation: superposition of the primary and mirror images.*
  c) *Noise-activated afterimages after a monocular left-eye fixation. The reversals occur for each letter and for their order through the right eye.*
  d) *Similar results after a right-eye fixation.*

complete word like NEURONS in Fig. 8a, CT perceived the primary image of the whole word through one eyelid (Fig. 8b, c) but also the whole mirror-image alone through the eyelid of the eye which has remained closed (Fig. 8b, c). Curiously, here CT sees through this eye "the words

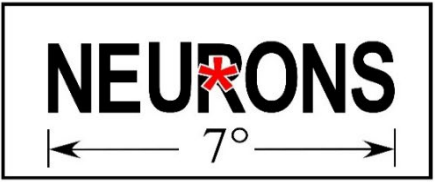

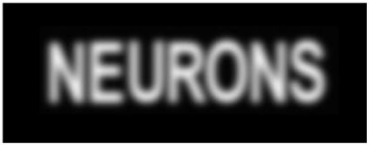
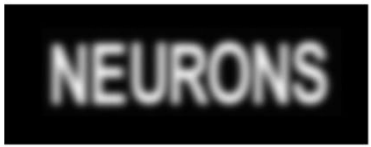

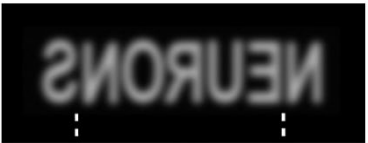
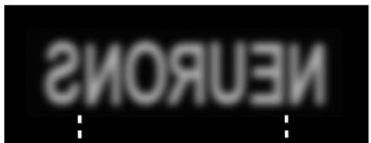

*Figure 8. Complete word afterimages.*
   a) *Stimulus: NEURONS.*
   b) *After a left-eye fixation the noise-activated primary afterimage is seen in the left eye and the whole reversed mirror image is seen in the right eye. The letters and the whole word are reversed.*
   c) *Similar results after a right-eye fixation.*

---

going the wrong way through the looking glass" like a Lewis Carroll character, contrary to the typical reader (see Supplementary S2). More surprisingly, a reader with mirror-images is also able to read reversed mirror non-words. Although we can try to decipher the reversed non-words (Fig. 9a) one letter at a time, we are rather helpless when faced with such an exercise. In contrast, after a monocular fixation, due to the noise, CT is able to read directly through the eyelid of the eye which has remained closed, the non-word alone with its restored orientation on the dominance columns of the layer 4 (Fig. 9b, c), contrary to the typical reader (Supplementary S3).

a) Stimulus: inverted non-word

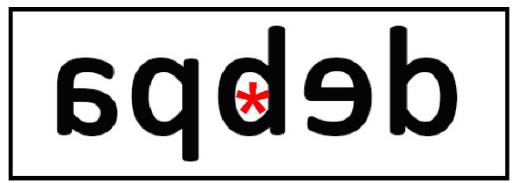

b) **Left-eye fixation**  c) **Right-eye fixation**

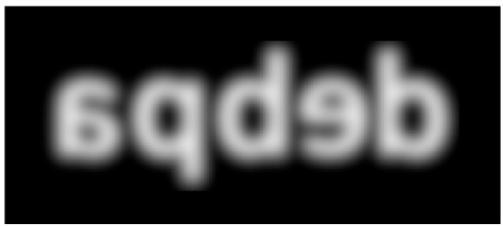

Left-eye afterimage

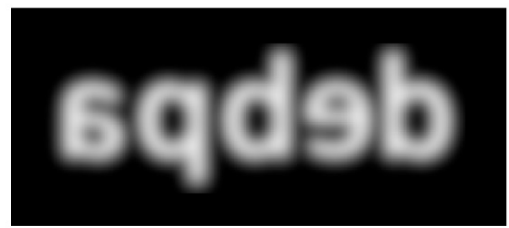

Right-eye afterimage

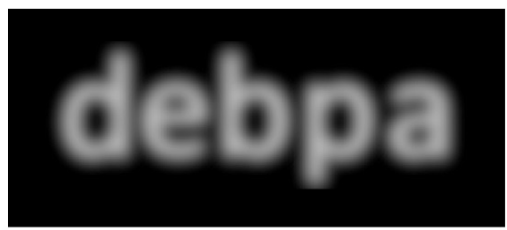

Right-eye afterimage

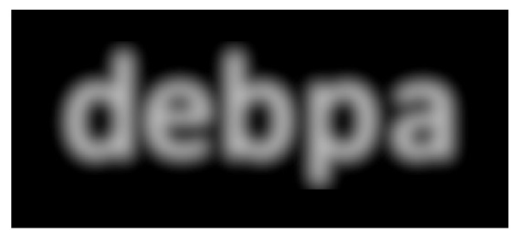

Left-eye afterimage

*Figure 9. Inverted non-word afterimages.*
  a) *Stimulus: inverted non-word.*
  b) *Left-eye fixation: inverted noise-activated afterimage in the left-eye, but restored orientation of the stimulus in the right-eye.*
  c) *Similar observations after a right-eye fixation.*

*5-The WAS and SAW confusion*

Let us consider the often cited tendency to read the word "WAS" as "SAW"[49] . The mechanism can be investigated using the noise-activated afterimage technique. After a binocular fixation of the stimulus "WAS" (Fig. 10a) in a continuous lighting, CT perceived simultaneously the veridical word 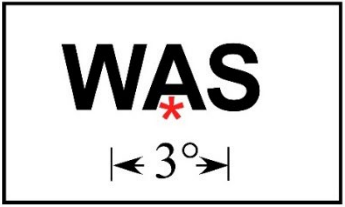 and the mirror-image ƨAW inducing the confusion (see Fig. 10b). However, the monocular fixation shows that while the primary image is seen

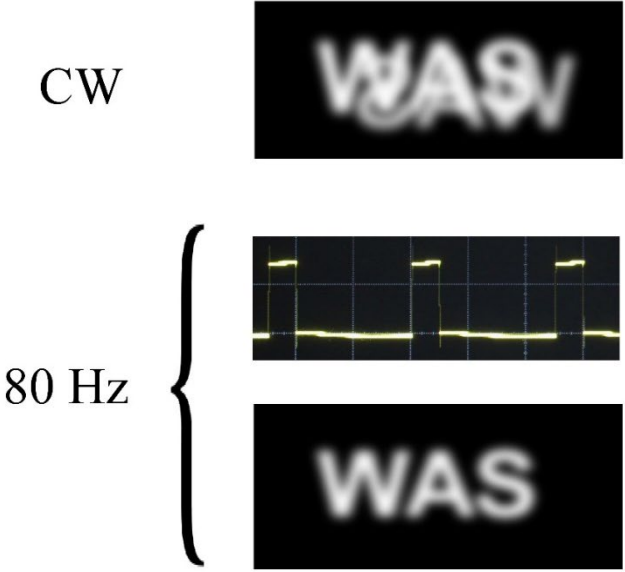

*Figure 10. Mirror-image erasing for the common mistake "WAS" and "SAW"* [49] *.*

a) *Stimulus: WAS*
b) *After binocular fixation: in continuous lighting (CW) both primary and mirror-images are seen superposed; in pulse lighting (80 Hz) only the primary image is seen, as for a typical reader.*

-------------------------------------------------------------------------------------

through the fixating eye eyelid, the mirror-image is seen also through the other eye which has remained closed. Moreover, here again the mechanism is not simply a permutation of letters, but a complete full reversal of the word and letters illustrating the internal visual crowding. However, repeating the test after a binocular fixation in a 80 Hz modulated lighting[17], exploiting the typical small 10 ms delay of the projections, CT observed that the annoying extra mirror-image disappears (Fig. 10b), the corresponding internal visual crowding being erased thanks to the Hebbian mechanisms[31].

## III- Discussion - Conclusion

In spite of the functional lateralisation of the human brain, the connections between the two hemispheres play a crucial role in its organization[22,50]. In vision in particular, due to the decussation at the chiasma for each eye, the callosum has to cement together the two halves of the visual world for each eye[12], ensuring continuity. Moreover, symmetric connections between the two hemispheres due to the mirror symmetry principle are specially abundant at birth and partially pruned after birth and during the first years of the critical period corresponding to the development and the maturation of the visual system[20,25,32,51,52], leading to mirror-images for young children and for many persons with dyslexia. By exploiting noise falling on the nonlinear bistable photoreceptors of the retinas, our results show the existence of mirror-images which are perceived and seen superposed on the primary afterimages observed after a binocular fixation.

The complete architecture of the neural connections that underlie symmetry perception can then be investigated with the help of a young observer with dyslexia and mirror-images. Indeed, after extending our noise-activated afterimage method to a monocular fixation, the superposition of primary and mirror-images through the closed eyelids is spatially resolved, as the primary image is perceived alone through the fixating eye, while the mirror-image is perceived also alone, but only through the other eye which has remained closed. We are forced to conclude that the noise activated afterimages occur on the dominance columns of layer 4 of V1, the only layer with a strict segregation between the signals coming from the two eyes [12,22,28]. Moreover, the projections through the callosum are necessarily crossed between the respective dominance columns of the two eyes. For instance, the primary image of b in Fig. 5, seen alone on the dominance columns of the left eye which has performed the monocular fixation, is also seen alone as the mirror image d, but through the right eye, implying that the mirror projections are crossed between the respective dominance columns of layer 4 of the two eyes (Fig. 6). Besides, although less frequently, duplicated interhemispheric projections of letters and words are also observed in persons with dyslexia[17,19]. Although less disturbing than mirror-images, these duplicated non-symmetric projections are also crossed between the layer 4 dominance columns of the two eyes. Our results show the crucial role played by the layer 4, thanks to its specific properties which are unique in the primary cortex. Transcriptomics has recently shown that layer 4 in V1 is clearly enlarged in humans[53]. First, most of the afferents from the ganglion cells of the retinas transiting through the lateral geniculate nucleus are projected on layer 4. Second, the information from each eye is preserved as layer 4 is the only layer where the segregation is strict[12]. Third, the layer 4 does not receive any feedback which could perturb its retinoptic map[28,54,55]. Fourth, the receptive fields of the neurons in layer 4 are the smallest in the cortex V1[56]. Moreover, layer 4 is the only layer that is insensitive to the orientation of stimulus[12], and so, in contrast to the other layers is sensitive to diffuse light, i.e. to the noise passing through the eyelid in our method. Hence, layer 4 appears to be ideal for receiving the precise encoded retinotopic maps[57], namely the mirror projections activated here by noise. In addition, the layer 4 contributes to the quality of the other retinoptic maps transmitted to the other layers observed by other authors[57,58]. Patterns of synaptic connections in the visual system are remarkably precise[59] and determine the quality of visual maps and visual perception. Moreover, layer 4 plays also an important role in the somatosensory cortex[60], and shows a developmental plasticity until the end of the visual critical period[61].

Still, one may wonder what is the real interest of the crossed architecture of the interhemispheric connections between the dominance columns of layer 4 across typical individuals. At first glance, the perceived interhemispheric projections that we observe here seem only to bring difficulties for a clear vision. Perceptual crowding by external flankers has previously been shown to represent an essential bottleneck for object perception and reading[62–65]. Here, we are confronted with mirror-images which produce a delayed internal visual crowding, in absence of any external distractors. Perception similarity of mirror-images has already been reported early in infancy[66], babies being unable to discriminate a 45° oblique from its mirror-image. Most children in all cultures tend to confuse left and right when they read and write[67], making mirror errors. Such confusions usually disappear in children around 5-6 years old, which corresponds to the end of the visual critical period[68], concomitantly with the consolidation of their ocular dominance[39]. Moreover, for young children it is only at around 6 years old that vision also becomes dominant over touch where *Piezo2* gene is the major transducer [69], namely for mirror-image discrimination[70], and later in spatial delocation which promotes largely visual dominance[71]. The end of the critical period appears as a turning point in the development and maturation of the visual system.

In typical adults, an asymmetry of the Maxwell's centroids (Fig. S1) corresponding to the blue cone-free area at the centre of the foveas has been shown to determine the ocular dominance, the quasi-circular profile corresponding to the dominant eye[17]. The topography of the cones in the fovea of the dominant eye being more regular than that of the non-dominant eye, the accuracy of the imprinted image on the retina of the dominant eye is slightly better in this eye. This small difference induces a symmetry breaking [72] between the two eyes leading to approximately two-thirds of observers being right-eye dominant and one-third being left-eye dominant[73]. The corresponding retinoptic maps of the primary images encoded in the ocular dominance columns for each eye in the layer 4 are also necessarily slightly different. The symmetric point-by-point transfer of the projections of mirror-images through the callosum being crossed, the retinal asymmetry is then transferred to the dominance columns of the layer 4 of the primary cortex. Indeed, crossed projections from the cortical topography in the dominance columns of one eye to a different topography in the columns of the other eye induces necessarily induces a slight degradation of their quality, leading to the weakening and suppression of the transferred mirror-images read by the brain, as we observe in typical individuals (Figs. S2, S3). Thus, during development and maturation, the differences linked to the asymmetry of the retinas emerge at the layer 4 level. In the Supplementary materials, we show the asymmetry of the Maxwell centroids for a left-dominant and a right-dominant adult, with only their respective primary afterimages after a monocular fixation. No mirror-images are observed in the eye which has remained closed (Figs. S2, S3).

However, to reach such an adult-like visual neural status, including lateralisation[74], requires a rather long development and maturation[75–79]. Interestingly, during childhood, children remain less foveally-biased compared to adults, their fixation on words being more left and upper-field biased than adults[80]. Although the blue cone-free areas on the retinas are adult-like as soon as 20 week gestation of pregnancy when the migration of the blue cones is important[33], each eye has to sculpt its pathways to the cortex and to the higher levels, particularly during the visual critical period where the plasticity is high, avoiding the possible neuronal migration anomalies[81]. During the development of the whole visual system, the retinal waves play an important role in the post-natal refinement of the visual circuits[82], in the segregation of the visual and somatory circuits[83] and in the modular strategy for the evolution of the hierarchical visual network[84]. Moreover a discrete genetically specified neuronal network instructing the neural circuit assembly has been recently discovered[85]. In addition, it has recently been shown that the end of the critical period is controlled by the astrocytes[86]. Curiously, the production of connexin 30 which closes the critical period for visual plasticity occurs precisely in the layer 4 where the projections take place, but only when in humans the mirror-images are no longer observed. However, although adults no longer perceive the mirror images as we observe in Figs. S2, S3, these images remain encoded as weak memory traces in the upper layers of the brain. The brain harbors visual representations in their normal and reflected forms, which can be selectively damaged. Indeed, following a traumatic brain injury, a patient with only acquired mirror reading and writing has been reported, the mirror-image representation alone having been preserved[87]. The lack of asymmetry in the topographies of the photoreceptors between the two retinas linked to the absence of ocular dominance, i.e. maintaining a competition between the two eyes, can also perturb the development of the connectivity of the magno, parvo and konio visual pathways and possibly plays a role in the functional lateralisation of the brain [88]. Our results appear compatible with the different impairments already clearly observed in many dyslexics, in particular in the frame of the visual magnocellular hypothesis[3,89,90]. Moreover, the competition can also influence the functional connectivity in the retina[91] and in the whole brain[92], the organization[93] and the sensory map topography in the primary visual cortex[94]. Beyond the spatial separation of the primary and projected images, the transit time through the callosum reaches about 10 ms[30]. Thus, the primary and mirror images appear spatially and temporally resolved leading to the possibility of weakening the embarrassing internal visual crowding in dyslexia. Using a pulsed light[17] at a frequency of about 80 Hz, we show that the Hebbian mechanisms[31] at synapses in layer 4 erase the projections are erased as shown in Fig. 10.

In conclusion, we show that binocular and monocular noise-activated afterimages seen by an observer with dyslexia and with mirror-images which lacks of asymmetry between his Maxwell centroids, shed light on the nature of the callosal interhemispheric connections in the human brain. First, binocular fixations confirm the superposed existence and perception of both primary and mirror images occurring often in dyslexia. Second, we deduce from the monocular fixations that the interhemispheric transfer of the mirror-images needs to occur on the layer 4 of the primary cortex, the only layer with a strict segregation between the two eyes, but moreover with a spatial crossing between the ocular dominance columns of the two eyes. Such an architecture suggests that for typical individuals, with an asymmetry between the Maxwell centroids, the crossed geometry also transfers this asymmetry to the layers 4 of the cortex, weakening the projected embarrassing mirror-images which are eliminated at the end of the critical period. Third, the mirror-images are perceived by the observer with dyslexia not only in reversal of individual letters, but also in complete reversal of letters in bigrams and in the reversal of whole words, leading to an internal visual crowding. However, as the persisting mirror-images require an extra travel time through the splenium of the callosum, they are slightly delayed and we are able to weaken these mirror-images via the Hebbian mechanisms using pulsed optical lighting. Paradoxically, our results suggest that with the help of visual noise, the observers with dyslexia disturbed by mirror-images, could provide new insights into the architecture and the dynamics of the callosal interhemispheric projections linked to the degree of functional lateralisation in the human brain[92,95,96], both for the observers with dyslexia and typical individuals.

## IV-Methods

*1-Participants.*

The different experiments were performed first by CT, a 18 years old student as an observer with dyslexia and mirror-images, and by two readers with normal binocular vision one with a right eyedness and a second with a left eyedness. Informed consensus was obtained for each observer. All experiments are non invasive. The entire investigation process was conducted according to the principles expressed in the Declaration of Helsinki.

*2-Noise activation of negative afterimages after binocular and monocular fixations.*

The method is based on the nonlinearity and the bistable dynamics of the photoreceptors of the retinas and of the layers of neurons along the visual pathway to the layer 4 of the primary cortex. When looking at a contrasted stimulus, a differential bleaching of the photoreceptors occurs, corresponding to the different parts of the stimulus. The sensitivity to an external noise after a fixation will depend on the relative bleach level of the photoreceptors. In eq. 3, in place of the noise $\xi(t)$, we can introduce an "effective noise" $k\,\xi(t)$, where $k \simeq 1$ for the dark unbleached parts of the stimulus and $k \simeq 0.5$ for the bright bleached parts. Such different bleach levels are reached after a fixation of 5 to 10 s when looking through a window in normal daylight with an illuminance of about 7000 lux. In all the experiments the illuminances are measured using a Roline RO-1332 digital flux meter. The black letters and words taped on the window have typically a width of 2 to 3 cm. The fixation is performed at 2 m from the window, where the illuminance is about 1600 lux. As a closed eyelid has a transmittance of about 2%, in the experiment, the noise reaching the corneas corresponds to an illuminance of diffuse light of about 30 lux, leading to contrasted and precise afterimages on the layer 4 of the primary cortex.

In the binocular fixation, the observer fixates the letters of the word during 10 s, closes his eyes and blocks any light impinging on his eyelids with his hands. Second, modulating the noise passing through his eyelids by shifting his two hands in front of his eyes with a periodicity of about 2 s, he activates the negative afterimages each time the noise reaches the cornea. As the fading time of the negative afterimages is extended to tens of seconds when the noise is modulated, the observer can observe and read the afterimages about ten times successively. For words the subtended angular diameter can reach 7°, i.e. the afterimages can exceed the limits of the fovea. For the dyslexic person with mirror-images, the primary and the mirror images are then both perceived and their coexistence produces an internal visual crowding.

In the monocular fixation, the observer also fixes the letter or the word during 10 s through either the right or the left eye. After blocking the noise in each eye with the two hands, he shifts with a period of about 2 s the hand from the eye that has made the fixation, then alternates by shifting the hand in front of the eye which has remained closed during the fixation. A typical observer sees the primary afterimage through the eyelid of the fixating eye, and nothing through the other eye. In contrast, the observer with dyslexia sees the primary image alone through the eyelid of the fixating eye as the typical observer, but sees the mirror-image also alone through the second eye which has remained closed during the fixation. So the monocular fixation method permits to spatially and temporally separate the two overlapping afterimages as the mirror-image is seen on the crossed dominance columns of the other eye and so has to travel further through the callosum which takes about 10 ms[30]. Changing the fixating eye gives the same results and shows that the neural connections remain unchanged.

*3-Maxwell centroids recordings*

The only direct observations of the 100 to 150 microns diameter blue cone-free areas at the centre of the foveas have been performed post-mortem[33,35]. Collecting intact retinas and staining the blue cones with antibodies is specially delicate, hence comparing the two outlines of the blue cone-free areas of the two eyes for a given patient was out of reach. To try to measure a possible asymmetry between the two eyes of any observer which would break the symmetry as expected[1,12], we use a foveascope able to visualise and enlarge the blue cone-free area images corresponding to the Maxwell centroids, i.e. the small central part of the whole

Maxwell spot observable in the blue part of the spectrum. To obtain a contrasted entoptic image of the Maxwell spot we have to brightly illuminate a screen with a projector of 3000 Lumens, giving a 4500 Lux illuminance on the screen located at a distance of 1.2 m. The observer looks at the bright screen through a blue-green exchange filter which alternates the two colours with a periodicity of about 7 s taking account of the fading time of 6-8 s of the entoptic image. For a distance of 3 m between the observer and the screen , the mean diameter of a typical Maxwell centroid image is about 3 cm on the screen. The transmissions of the optimised two colour filter we use are shown in Fig. S4. Associating a green filter to a blue filter to have a good contrast is suggested by the fact that observers who lack green cones, i.e. the deuteranopes are unable to see their Maxwell spots[97]. For a typical reader the centroid for one eye shows a quasi-circular outline for one eye, the dominant eye, and an elliptical outline for the non dominant eye (see Figs. S2 and S3). In contrast, for a person with dyslexia the two outlines are generally quasi–circular ( Fig. 1b) without any appreciable asymmetry, leading usually to no ocular dominance. First, for each eye, an observer compares the outline of his blue cone-free area with calibrated circles projected on the screen to measure the mean angular diameter of the Maxwell centroid. Second, when the outline of the entoptic image is elliptic, namely for the non dominant eye, adjustable ellipses are projected onto the screen with variable size, ellipticity and inclination $\varphi$ of the main axis (see Fig. S1). The observer optimises the superposition of the entoptic image outline and the projected ellipse. Hence, for each observer the asymmetry between the two eyes can be completely specified.

**Acknowledgments**

We thank the University of Rennes for access to its facilities, and J. R. Thébault for his technical assistance. The authors thank the observers for their kind participation and K. Dunseath, B. Mitchell, and R. Le Naour for reading the manuscript and for their comments.

**Authors contributions**

ALF designed the study and wrote the manuscript. GR and ALF developed the methods and the apparatus and contributed to the interpretation of the data and discussed the results.

**Competing interest**

Patents have been filed by the university of Rennes. A partnership has been signed between the university of Rennes and the company ATOL-Abeye and the company Lili for Life.

**Additional information**

Supplementary material are available.

# Supplementary Information: Direct observation of the crossed interhemispheric transfer of the left-right mirror-images in human vision

Albert Le Floch [a,b], Guy Ropars [a,c*]

[a]*Laser Physics Laboratory, University of Rennes, 35042 Rennes Cedex, France*

[b]*Quantum Electronics and Chiralities Laboratory, 20 Square Marcel Bouget, 35700 Rennes, France*

[c]*UFR SPM, University of Rennes, 35042 Rennes Cedex, France*

*\* Corresponding author at: Laser Physics Laboratory, University of Rennes, 35042 Rennes Cedex, France. e-mail: guy.ropars@univ-rennes.fr ; albert.lefloch@laposte.net*

The crossed nature of the projections between the dominance columns of the layer 4 described in the context of the observer with dyslexia, also plays a crucial role in normal individuals, since until the end of the visual critical period at about 5-6 years old the young children generally confuse the letters with their mirror-images. The end of the critical period is a turning point when the mirror-images disappear concomitantly with the stabilization of the ocular dominance[36,37]. Here, using the foveascope, we have recorded the profiles of the Maxwell centroids for a right-eye dominant and a left-eye dominant adult (Fig. S1). The asymmetry defined thanks to the osculating ellipses for the two Maxwell profiles is equal to $\Delta\varepsilon = \varepsilon_R - \varepsilon_L = + 0.4$ for the right-eye dominant observer and to $- 0.5$ for the left-eye dominant observer.

In Figs. S2 and S3, the two normal observers have recorded their noise activated afterimages perceived in each eye using the monocular fixation method. While the primary image is perceived with the eye which performs the fixation, no mirror-afterimage is perceived in the other eye. The crossed point-to-point symmetric callosal projections between the dominance columns of the two eyes with their different neuronal topographies resulting from the asymmetry have been weakened and the mirror-images are no longer perceived. To optimize the contrast of the Maxwell centroids the different useful functions are shown in Fig. S4.

## a) Right-eyed reader

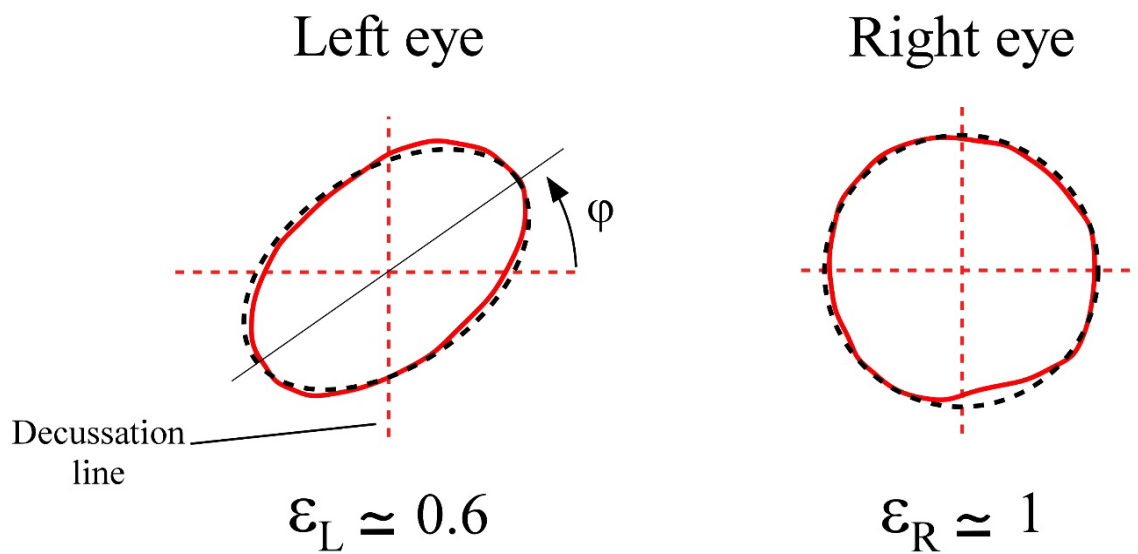

$\varepsilon_L \simeq 0.6$    $\varepsilon_R \simeq 1$

## b) Left-eyed reader

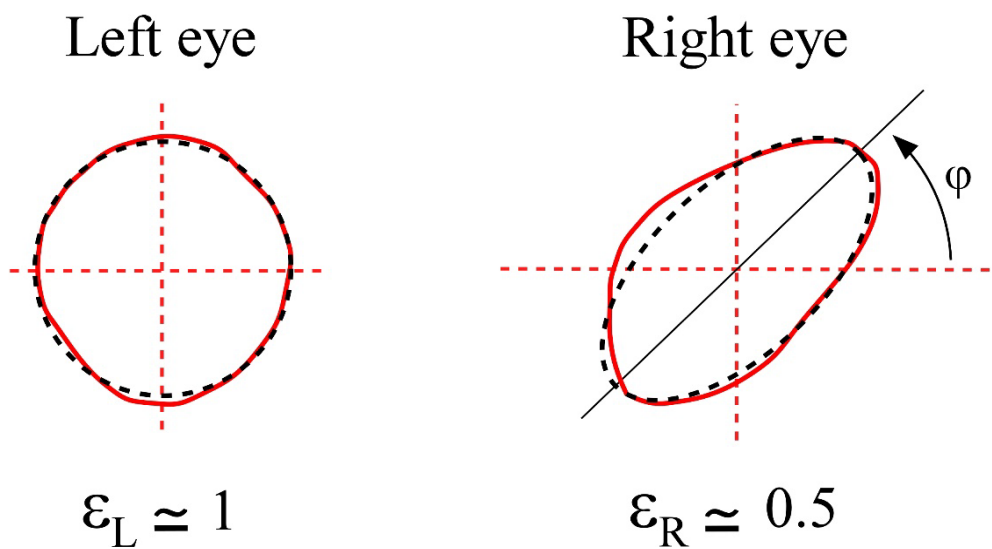

$\varepsilon_L \simeq 1$    $\varepsilon_R \simeq 0.5$

**Supplementary Figure S1**: Maxwell centroid outlines recorded by two normal readers.

The normal readers with a right eye and a left eye dominance respectively, show opposed asymmetry and ocular dominance. The angle $\varphi$ determines the orientation of the ellipse.

## a) Stimulus: word

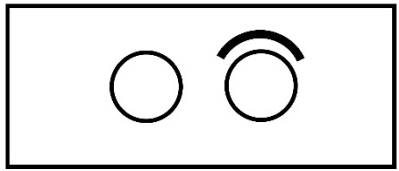

## b) **Monocular left-eye fixation by a normal reader**

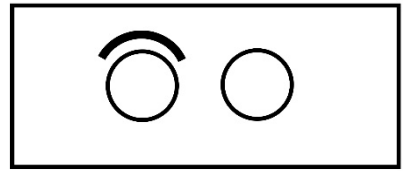

## e) **Monocular right-eye fixation by a normal reader**

## c) Left-eye afterimage

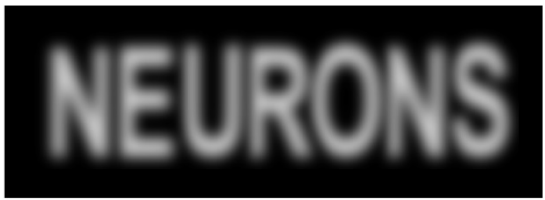

## f) Right-eye afterimage

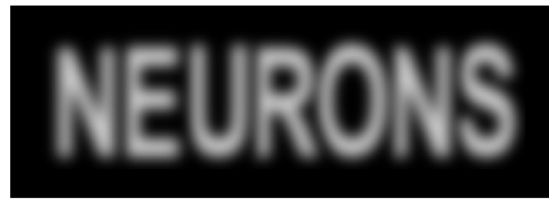

## d) Right-eye afterimage

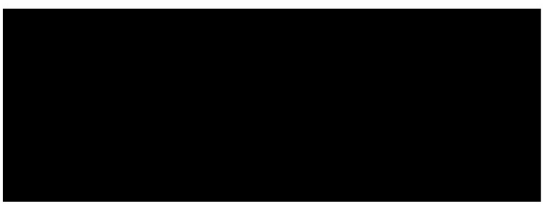

## g) Left-eye afterimage

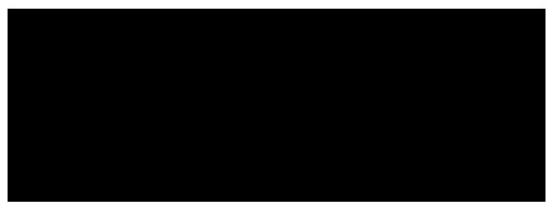

**Supplementary Figure S2**: Noise activation after a monocular fixation for a word for the two normal readers.
No mirror-image is perceived on the dominance columns on the eye which had remained closed.

## a) Stimulus: inverted non-word

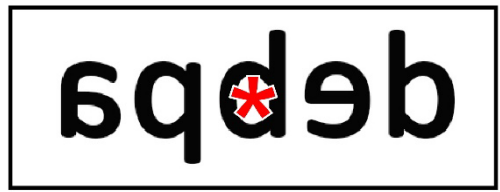

b) **Monocular left-eye fixation by a normal reader**

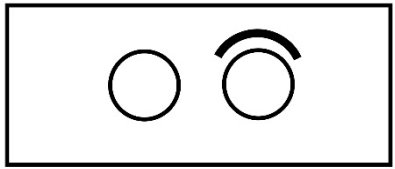

e) **Monocular right-eye fixation by a normal reader**

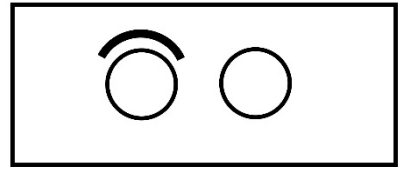

c) Left-eye afterimage

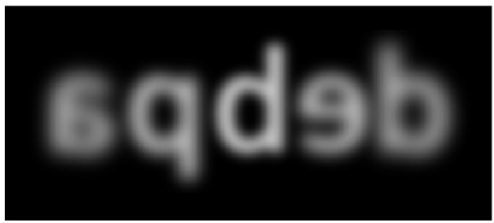

f) Right-eye afterimage

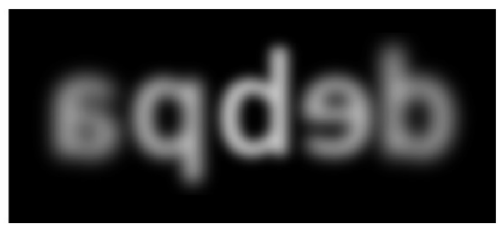

d) Right-eye afterimage

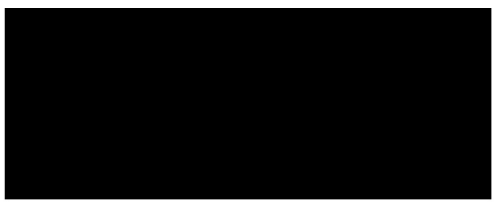

g) Left-eye afterimage

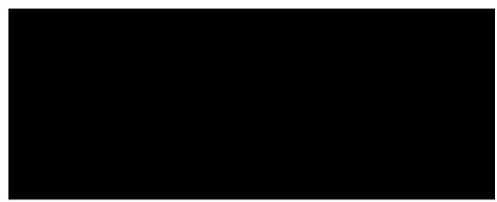

**Supplementary Figure S3**: Noise activated afterimages after monocular fixation. Observations for the two normal readers for a reversed non-word. In contrast to CT (see Fig. 9), they are left with only the primary afterimage corresponding to the reversed stimulus.

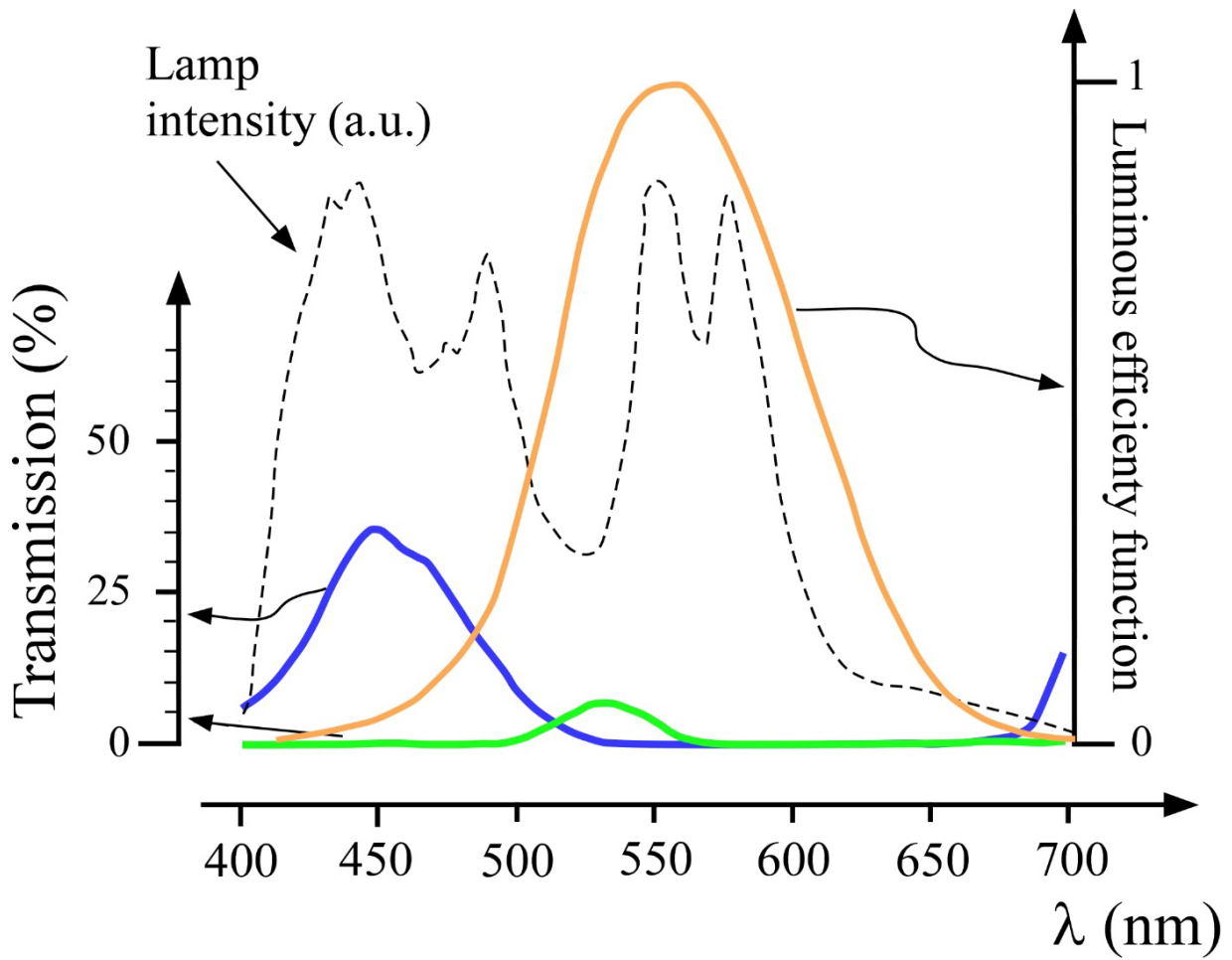

**Supplementary Figure S4**: Respective transmissions of the blue and green parts of the exchange filter used to record the Maxwell centroid outlines, with the lamp intensity projected on the screen (dotted line), and the luminous efficiency function of the human eye (orange line). The transmissions have been optimized so as to produce the impression of same brightness through the blue and green filters.